\documentclass[fleqn,usenatbib]{mnras}

\usepackage{newtxtext,newtxmath}

\usepackage[T1]{fontenc}
\usepackage{orcidlink}
\DeclareRobustCommand{\VAN}[3]{#2}
\let\VANthebibliography\thebibliography
\def\thebibliography{\DeclareRobustCommand{\VAN}[3]{##3}\VANthebibliography}

\usepackage{graphicx}	
\usepackage{amsmath}	
\usepackage{bm}

\title[Mergers driven by circumbinary discs]{Mergers of black hole binaries driven by misaligned circumbinary discs}

\author[R. G. Martin et al.]{
Rebecca G. Martin$^{1,2}$\thanks{E-mail: rebecca.martin@unlv.edu}\orcidlink{0000-0003-2401-7168},
Stephen Lepp$^{1,2}$\orcidlink{0000-0003-2270-1310},
Bing Zhang$^{1,2}$\orcidlink{0000-0002-9725-2524},
C. J. Nixon$^3$\orcidlink{0000-0002-2137-4146}
and 
Anna C. Childs$^4$\orcidlink{0000-0002-9343-8612}
\\
$^1$Nevada Center for Astrophysics, University of Nevada, Las Vegas,
4505 South Maryland Parkway, Las Vegas, NV 89154, USA\\
$^2$Department of Physics and Astronomy, University of Nevada, Las Vegas,
4505 South Maryland Parkway, Las Vegas, NV 89154, USA\\
$^3$School of Physics and Astronomy, Sir William Henry Bragg Building, Woodhouse Ln., University of Leeds, Leeds LS2 9JT, UK\\
$^4$Center for Interdisciplinary Exploration and Research in Astrophysics (CIERA) and Department of Physics and Astronomy Northwestern University,\\ 1800 Sherman
Ave., Evanston, IL 60201, USA
}

\date{Accepted XXX. Received YYY; in original form ZZZ}

\pubyear{2023}

\begin{document}
\label{firstpage}
\pagerange{\pageref{firstpage}--\pageref{lastpage}}
\maketitle

\begin{abstract}
With hydrodynamical simulations we examine the evolution of  a highly misaligned circumbinary disc around a black hole binary including the effects of general relativity. We show that a disc  mass of just a few percent of the binary mass  can significantly increase the binary eccentricity through von-Zeipel--Kozai--Lidov (ZKL) like oscillations provided that the disc lifetime is longer than the ZKL oscillation timescale. The disc begins as a relatively narrow ring of material far from the binary and spreads radially.  When the binary becomes highly eccentric, disc breaking forms an inner disc ring that quickly aligns to polar. The polar ring drives fast retrograde apsidal precession of the binary that weakens the ZKL effect. This allows the binary eccentricity  to remain at a high level and may significantly shorten the black hole merger time.  The mechanism requires the initial disc inclination relative to the binary to be closer to retrograde than to prograde.
\end{abstract}




\section{Introduction}

A body orbiting a black hole binary in an inclined orbit can cause oscillations of the binary eccentricity \citep{Farago2010,Aly2015,Naoz2016}. These von-Zeipel--Kozai--Lidov oscillations \citep[ZKL,][]{vonZeipel1910,Kozai1962,Lidov1962} can occur even in the case that the outer body is much less massive than the inner binary \citep[][]{Chen2019,Hamers2021}.  Increased  eccentricity  drives the merger of the black holes more rapidly as a result of energy loss from gravitational radiation \citep{Peters1964,Antonini2012,Silsbee2017}. For a very high mass third body, the maximum eccentricity is achieved when the third body is in a polar orbit relative to the binary orbit \citep{Liu2018,Liu2019,Liu2019b}. In this configuration,  the angular momentum vector of the third body orbit is aligned to the eccentricity vector of the binary.  However, recently we found that for a lower mass third body,  a mass of just a few percent of the binary mass, the largest binary eccentricity growth occurs for inclinations that are initially closer to retrograde than to prograde \citep{Lepp2023b}.

A black hole binary may accrete material in a chaotic fashion at random inclinations relative to the binary orbit. This is how supermassive black holes at the centres of galaxies are thought to grow \citep{King2006,Nixonetal2012a,King2015}. Similarly chaotic accretion on to young binary stars is seen in simulations of star forming regions  \citep{Offner2010,Bate2012,Bate2018} and observations show that misaligned circumbinary discs are common \citep[e.g.][]{Chiang2004,Kohler2011,Brinch2016,Czekala2019}.

The orbit of a massive body around a binary can undergo complicated behaviour relative to the binary orbit \citep{Verrier2009,Farago2010,Doolin2011,Naoz2016,Chen2019}. A misaligned orbit undergoes nodal precession and the binary eccentricity oscillates on the same timescale. The nodal precession may be centered on the binary angular momentum vector for low initial inclination (circulating orbits) or centered on a stationary inclination for higher initial inclination (librating orbits).  For a test particle, without the effects of general relativity (GR), the stationary inclination is at $90^\circ$, a polar orbit in which the angular momentum vector of the outer body is aligned to the eccentricity vector of the inner binary. GR drives prograde apsidal precession of the binary \citep[e.g.][]{Ford2000,Naoz2017,Zanardi2018} that increases the stationary inclination  \citep{Lepp2022,Zanardi2023} while the mass of the  particle  lowers the stationary inclination \citep{Chen2019,Martin2019polar}. 
The maximum binary eccentricity is achieved when the third object is in a librating orbit \citep{Chen2019}.

At each radius in the circumbinary disc, the material  feels the same torque as a point mass at that radius. However, the rings of a disc communicate with each other in a wave-like manner (if the disc aspect ratio, $H/R$, is larger than the \cite{SS1973} viscosity parameter, $\alpha$) or through viscosity (if $H/R\ll \alpha$). If the communication is faster than the global precession timescale, then the disc can precess as a solid body \citep[e.g.][]{Lubow2018} otherwise the disc can become warped or even break \citep{Larwoodetal1996,Nixon2013,Nixon2016}.  Two parts of a broken disc can undergo nodal precession independently on different timescales. The evolution of the third body  depends upon its mass and its orbital radius. Therefore, the evolution of a circumbinary disc depends upon the distribution of mass within the disc and breaking may lead to significantly different behaviour.

In this work, we show that a massive circumbinary disc can drive ZKL  like oscillations of the binary eccentricity of an equal mass binary. This mechanism has been previously seen in the case of an extreme mass ratio binary composed of a planet and a star \citep{Terquem2010}. In that case, the eccentricity of the planet orbit oscillates.  In Section~\ref{simulation} we consider hydrodynamic simulations of a circumbinary disc around an equal mass black hole binary and $n$-body simulations for comparison. We show that a circumbinary disc with a mass of only a few percent of the binary mass can significantly increase the binary eccentricity and speed up the merger time. We find that disc breaking can lead to a highly eccentric binary for a long timescale rather than an oscillating eccentricity as is the case if the outer body is a particle rather than a disc \citep{Lepp2023b}. In Section~\ref{massive} we consider the application to a range of black hole masses and we draw our conclusions in Section~\ref{concs}.

\section{Circumbinary disc and particle comparison}
\label{simulation}

We consider a binary  composed of two equal mass black holes with masses $M_1=M_2=30\,\rm M_\odot$ and total mass $M_{\rm b}=M_1+M_2$. They are initially in an orbit with semi-major axis of $a_{\rm b}=10\,\rm R_\odot$ \citep[e.g.][]{deMink2017,Martin2018bh} with eccentricity $e_{\rm b}=0.2$.  In this section we first describe the initial conditions for the circumbinary disc hydrodynamic simulation. We then show an equivalent massive particle $n$-body simulation and then we compare this to the disc simulation. Finally we consider a four-body simulation for comparison to the circumbinary disc simulation in which the disc breaks into two components.

\subsection{Circumbinary disc set-up}
\label{discsetup}

\begin{figure*}
\begin{center}
\includegraphics[width=0.9\columnwidth]{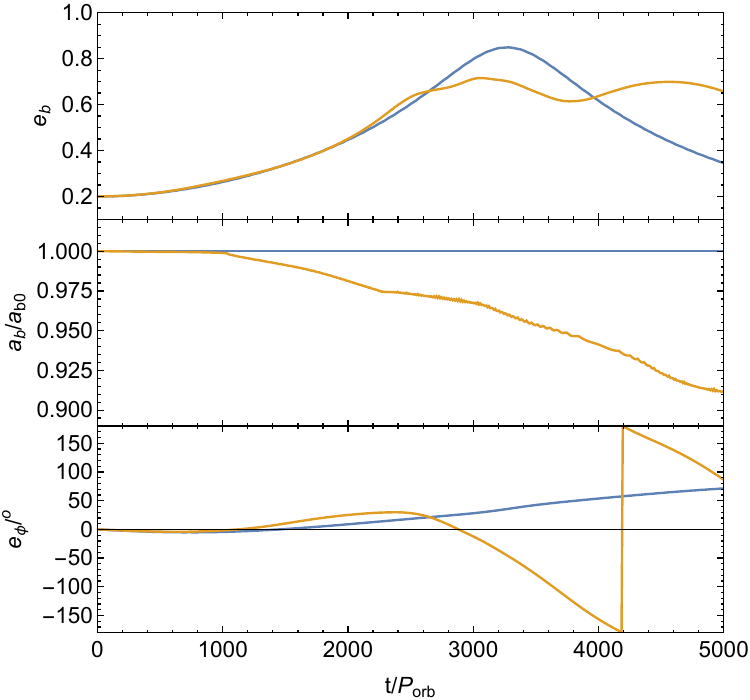}
\includegraphics[width=0.9\columnwidth]{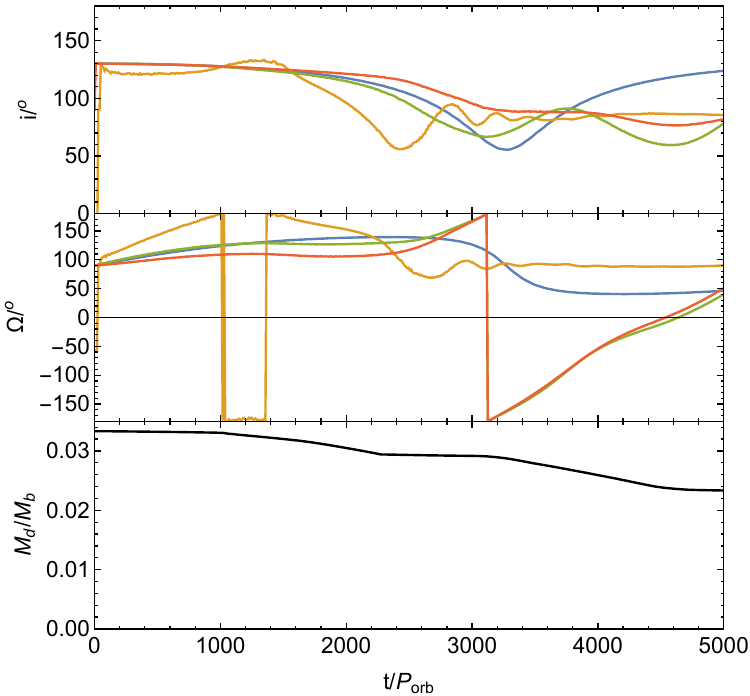}
	\end{center}
    \caption{Comparison of the circumbinary disc simulation to a three-body simulation. Left: The binary eccentricity (upper panel), binary semi-major axis scaled to the initial semi-major axis (middle panel) and phase angle of the binary eccentricity vector (lower panel) as a function of time for the  $n$-body simulation (blue) and the disc simulation (yellow). 
    Right: The inclination (upper panel) and longitude of ascending node (middle panel) of the third body (blue) and disc (at radius $R=4\,a_{\rm b}$ in yellow, $R=7\,a_{\rm b}$ in green and $R=10\,a_{\rm b}$ in red)  in the frame of the binary orbit. The lower panel shows the total mass of the disc.
     }
    \label{ecc}
\end{figure*}

We use the smoothed particle hydrodynamics \citep[SPH][]{Bateetal1995} code {\sc Phantom} \citep{Price2010,Price2018} to model the evolution of a circumbinary disc. This code has been extensively tested for circumbinary discs \citep{Nixonetal2012a,Facchinietal2013,Aly2018,Cuello2019,Abod2022,Smallwood2019}. The binary is modelled with sink particles that accrete the mass and angular momentum of any SPH particles that move inside their sink radius \citep{Bateetal1995}. The sink radius for each black hole is $2.5\,\rm R_\odot$. We include a modification to the sink-sink acceleration to model the effects of GR \citep{Nelson2000,Childs2023b} that drives prograde apsidal precession of the binary orbit \citep[e.g.][]{Zanardi2018}. 

The circumbinary disc is initially relatively far from the inner tidal truncation radius around the binary that is around $2-3\,a_{\rm b}$ for a coplanar disc \citep{Artymowicz1994} but smaller for a polar disc \citep{Franchini2019inner}. The initial total disc mass is $M_{\rm d}=0.033\, M_{\rm b}$ and is composed of $500,000$ particles that are distributed in a relatively narrow ring with surface density profile $\Sigma \propto R^{-3/2}$ between $R_{\rm in}=6.5\,a_{\rm b}$ and $R_{\rm out}=10\,a_{\rm b}$. The disc is free to spread inwards and outwards. 
The disc is initially inclined to the binary orbit by $i=130^\circ$ with longitude of ascending node relative to the binary eccentricity vector of $\Omega=90^\circ$. We also considered larger initial tilts, of $170^\circ$ and $150^\circ$ since these can drive even larger eccentricities for a particle \citep{Lepp2023b}. However, we did not find significant eccentricity growth with $i=170^\circ$. With $i=150^\circ$ we find similar behaviour to $i=130^\circ$ with only slightly lower eccentricity growth. We discuss this further in Section~\ref{massive}.

The disc is locally isothermal with sound speed $c_{\rm s}\propto R^{-3/4}$ and disc aspect ratio  $H/R=0.01$ at $R=6\,a_{\rm b}$. This choice allows the \cite{SS1973} viscosity parameter $\alpha$ and the shell-averaged smoothing length per scale height $\left<h\right>/H$ to be constant over the disc. We take the viscosity parameter to be $\alpha=0.1$, typical for fully ionised discs \citep[e.g.][]{Martin2019alpha}. The viscosity is implemented by adapting the SPH numerical viscosity with $\alpha_{\rm AV}=1.35$ and $\beta_{\rm AV}=2$ \citep{Lodato2010}. The disc is resolved with  $\left<h\right>/H\approx 0.7$ initially.

\subsection{Three-body simulation}
\label{three}

For comparison to the disc evolution, we first consider the evolution of a massive particle orbiting the binary. We use the $n$-body code {\sc Rebound} \citep{Rein2012} to model the evolution of the three-body system.

The angular momentum of the disc described in the previous section is
\begin{equation}
    J_{\rm d}=\int_{R_{\rm in}}^{R_{\rm out}} 2\pi R (R^2\Omega_{\rm K}) \Sigma \, dR,
\end{equation}
where the Keplerian angular velocity in the disc is
\begin{equation}
    \Omega_{\rm K}=\sqrt{\frac{G M_{\rm b}}{R^3}}.
\end{equation}
The angular momentum of a circumbinary particle of mass $M_3$ orbiting at semi-major axis $a_3$ is
\begin{equation}
    J_{3}=M_3 a_3^2 \Omega_3
\end{equation}
where
\begin{equation}
    \Omega_3=\sqrt{\frac{G (M_{\rm b}+M_3)}{a_3^3}}.
\end{equation}
We solve $J_{\rm d}=J_{\rm 3}$ to find that a particle with the same mass and angular momentum of the initial disc has semi-major axis $a_3=7.8\,a_{\rm b}$. In the particle simulation we incline the particle orbit by $i=130^\circ$
 and take $\Omega=90^\circ$ initially.

The inclination of the particle relative to the binary is
\begin{equation}
    i = \cos^{-1} \left( \hat{\bm{l}}_{\rm b}\cdot \hat{\bm{l}}_{\rm p}\right),
\end{equation}
where $\hat{\bm{l}}_{\rm p}$ is a unit vector parallel to the particle angular momentum vector and $\hat{\bm{l}}_{\rm b}$ is a unit vector parallel to the binary angular momentum vector. 
The longitude of ascending node is
\begin{equation}
  \Omega= \tan^{-1}\left(\frac{ \hat{\bm{l}}_{\rm p} \cdot (\hat{\bm{l}}_{\rm b} \times \hat{\bm{e}}_{\rm b})}{\hat{\bm{l}}_{\rm p}\cdot \hat{\bm{e}}_{\rm b}}\right)  +\frac{\pi}{2},
\end{equation}
where $\hat{\bm{e}}_{\rm b}=(e_x,e_y,e_z)$ is the binary eccentricity vector \citep[e.g.][]{Chen2020}.  The phase angle of the eccentricity vector of the binary in the plane of the initial binary orbit is calculated with
\begin{equation}
    e_\phi=\tan^{-1}\left(\frac{e_y}{e_x}\right).
\end{equation}

The left side of Fig.~\ref{ecc} shows the evolution of the binary eccentricity, semi-major axis and $e_\phi$ in time for the particle simulation in blue. The upper two panels of the right side of Fig.~\ref{ecc} show the inclination and  nodal phase angle of the third body relative to the binary for the three-body simulation in blue.  The eccentricity of the binary oscillates up to a maximum of 0.85 while the inclination oscillates to lower values as a result of the ZKL effect. The binary separation remains constant.

\subsection{Circumbinary disc simulation}
\label{sec:cbd}

\begin{figure}
    \centering
    \includegraphics[width=\columnwidth]{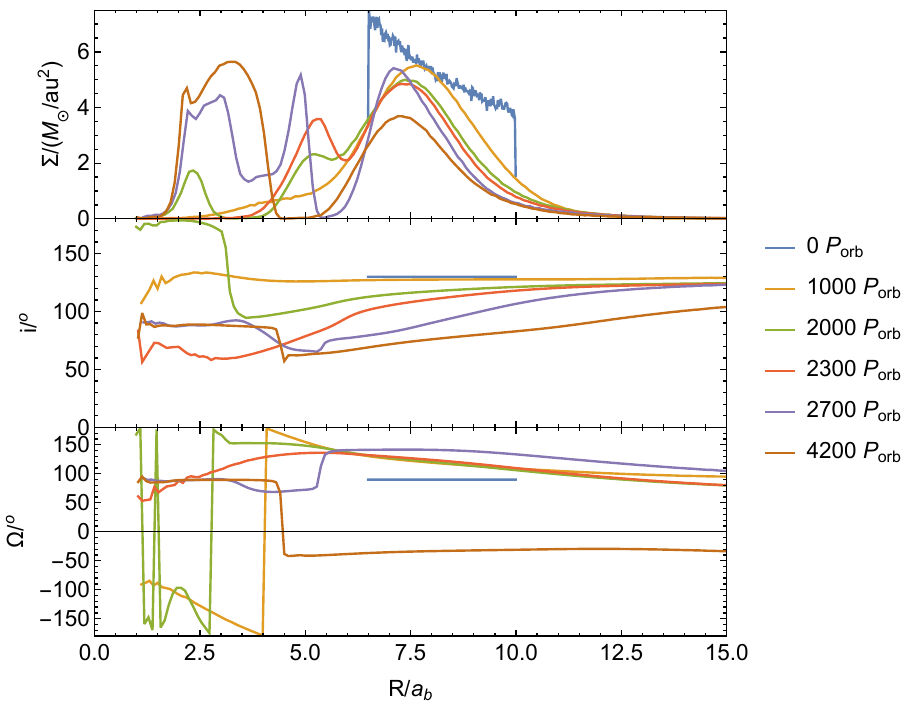}
    \caption{Surface density, inclination and longitude of ascending node as a function of radius at the same times as shown in Fig.~\ref{splash}.}
    \label{fig:sigma}
\end{figure}

The yellow lines in the left panel of  Fig.~\ref{ecc} show the binary eccentricity, semi-major axis  and $e_\phi$ in the circumbinary disc simulation. Initially the binary evolves in a very similar way to that in the particle simulation. However, as the disc evolves from its initial surface density distribution,  the behaviour begins to differ.

The upper two panels of the right panel of Fig.~\ref{ecc} shows the disc evolution at three different radii. The middle of the disc (green line) is initially very similar to the particle. However, the inner parts of the disc (yellow) show that the inner disc undergoes nodal libration and aligns to polar ($i=\Omega=90^\circ$).  The later evolution of the outer parts of the disc show a misaligned and circulating disc since the longitude of ascending node varies over $360^\circ$. The binary and the inner polar disc precess together with the combined effect of GR and the massive inner disc. The binary semi-major axis decreases in the disc simulation as a result of the accretion of low angular-momentum material from the disc. The times of high accretion rate (where the disc mass decreases more rapidly in the lower right panel of Fig.~\ref{ecc}) correspond to the times where the binary orbit shrinks more rapidly (middle left panel of Fig.~\ref{ecc}).

\begin{figure}
\begin{center}
\includegraphics[width=1.0\columnwidth]{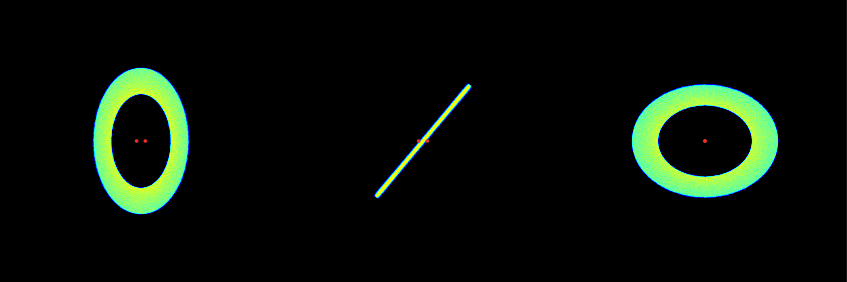}
\includegraphics[width=1.0\columnwidth]{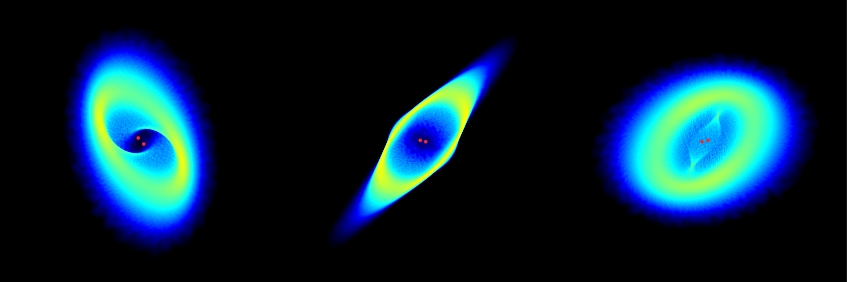}
\includegraphics[width=1.0\columnwidth]{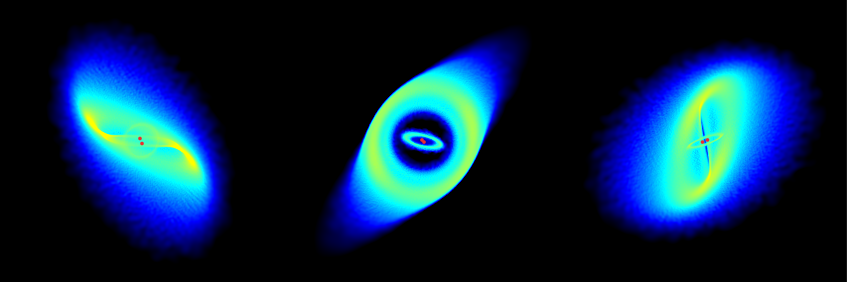}
\includegraphics[width=1.0\columnwidth]{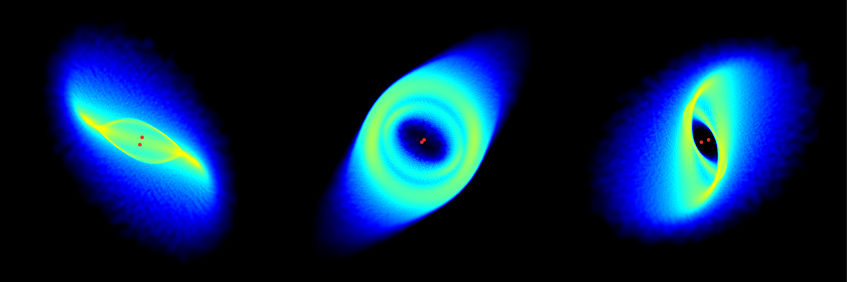}
\includegraphics[width=1.0\columnwidth]{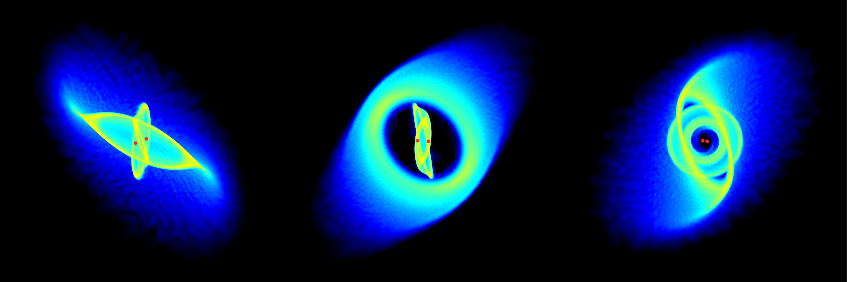}
\includegraphics[width=1.0\columnwidth]{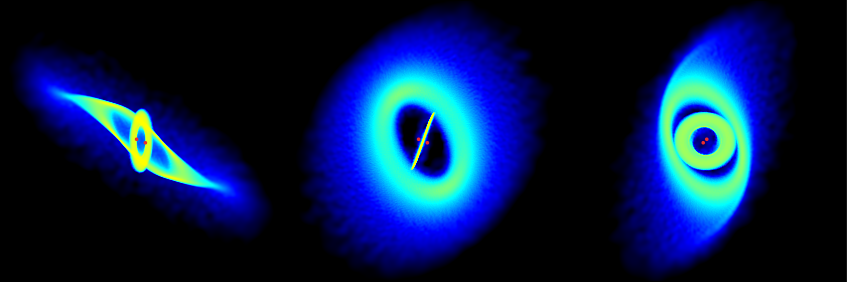}
	\end{center}
    \caption{Each row shows three views of the same disc: the disc in the $x-y$ plane in which the binary orbits initially (left), the $x-z$ plane (middle) and the $y-z$ plane (right). The times shown from top to bottom are $t=0$, 1000, 2000, 2300,  2700 and $4200\,P_{\rm orb}$. The red circles show the binary. 
    }
    \label{splash}
\end{figure}

Fig.~\ref{fig:sigma} shows the surface density, inclination and longitude of ascending node of the disc as a function of radius at different times.  Fig.~\ref{splash} shows the disc from three different viewing locations at the same times as the lines in Fig.~\ref{fig:sigma}. 
The top row in Fig.~\ref{splash} shows the initial setup. In the second row, at a time of $t=1000\,P_{\rm orb}$, the gas has flowed inwards towards the binary and is strongly warped close to the binary. At a time of $t=2000\,P_{\rm orb}$, a narrow inner ring has broken off inside a radius of about $2.6\,a_{\rm b}$ and aligned to retrograde (green line in Fig.~\ref{fig:sigma}).  This ring is accreted relatively quickly \citep[see also][]{Nixonetal2011a}.  At at time of $t=2300\,P_{\rm orb}$ the disc breaks again, this time farther out at about $R=6\,a_{\rm b}$ because the binary by now is much more eccentric. This inner ring aligns to polar and remains there for the remainder of the simulation. Between about $t=2500\,P_{\rm orb}$ and $t=3000\,P_{\rm orb}$ there are three rings in the disc with the inner two being polar. At later times, the two inner rings join together and there is just one radially wide inner polar ring.

In order for the outer disc to drive ZKL oscillations of the binary, the driving of the apsidal precession of the binary must be dominated by the outer disc,   otherwise its effect on the binary eccentricity averages to zero \citep{Holman1997}. The lower left panel of Fig.~\ref{ecc} shows the apsidal precession of the binary in the frame of the initial binary orbit. In the absence of other effects,  GR drives prograde apsidal precession of the binary on a timescale  $t_{\rm GR}=25,000\,P_{\rm orb}$ for $e_{\rm b}=0.2$ and this decreases to  $t_{\rm GR}=17,000\,P_{\rm orb}$ for $e_{\rm b}=0.8$ \citep[see equation 27 in][]{Naoz2017}. The disc drives prograde apsidal precession for inclinations close to coplanar \citep[e.g.][]{Lepp2023} but faster retrograde apsidal precession for polar inclinations \citep[e.g.][]{Zhang2019,Childs2023}.
The apsidal precession observed in the simulation is initially slightly prograde. This is a combination of the small prograde GR  plus a small prograde component from the inclined circumbinary disc (that is inclined by $i\gtrsim 100^\circ$ for all radii). The retrograde inner ring drives prograde apsidal precession.  Once the binary becomes highly eccentric and the polar inner ring forms, the apsidal precession is much faster and retrograde. This retrograde precession prevents ZKL oscillations from continuing and the binary eccentricity remains high. 

A polar disc ring drives retrograde apsidal precession of the binary at a rate three times faster than the prograde apsidal precession driven by an equivalent prograde (or retrograde) ring \citep[e.g.][]{Zhang2019,Childs2023}. The higher the binary eccentricity, the more likely it is that a polar ring can form since the larger the range of inclinations for which the orbits are librating  \citep[e.g.][]{Verrier2009,Chen2019}. This suggests that the locking of the binary eccentricity through the formation of an inner polar ring is much more likely for high binary eccentricity.

We have also considered simulations with a larger disc aspect ratio. With $H/R=0.02$, similar behaviour is observed with slightly reduced eccentricity growth. The peak eccentricity is about 0.6 and after the disc breaks, it settles to about 0.4. With $H/R=0.05$, there is no disc breaking, but because of the much shorter viscous timescale, there is not significant eccentricity growth before the disc has largely dissipated.  In the case that the disc is in good radial communication and does not undergo disc breaking, then the system behaves more similarly to the case with a third body instead of a disc and the binary eccentricity oscillates. This could occur for a disc with a much smaller viscosity $\alpha$ parameter than used in this work.

The merger of the black hole binary is driven by gravitational radiation on an approximate timescale $\tau\propto a_{\rm b}^4(1-e_{\rm b}^2)^{7/2}$ \citep{Peters1964}. A decrease in the semi-major axis of the binary of about 10\% (as seen in our simulation) leads to a merger speed-up factor of about 1.5.  While the gas disc is present, the ZKL effect leads to increased binary eccentricity  that  can remain high even after the gas disc has dissipated. 
The increase in the binary eccentricity up to about 0.7 leads to merger speed-up factor of about 10.
Therefore the increased eccentricity provides a larger merger time speed-up factor than the accretion of material that directly shrinks the binary orbit. We note that larger speed-up factors can be achieved for a larger disc mass and wider binaries that drive higher eccentricities \citep{Lepp2023b}.

While we did not observe significant binary eccentricity growth in the simulation with $i=170^\circ$, it is useful to compare the amount of binary shrinkage in this simulation since retrograde circumbinary discs are already known to shrink the orbit of a binary as a result of the accretion of material with negative angular momentum \citep{Nixonetal2011a,Nixonetal2011b}. For $i=170^\circ$, the binary semi-major axis decreases at a steady rate of that is less than half the rate observed in our $i=130^\circ$ case. There is no disc breaking for $i=170^\circ$  and the disc mass decreases about a factor of two more slowly. This suggests that the binary shrinkage observed in the $i=130^\circ$ simulation is a result of the accretion of low angular momentum material.

\subsection{Four-body simulation}

\begin{figure}
\begin{center}
\includegraphics[width=1.0\columnwidth]{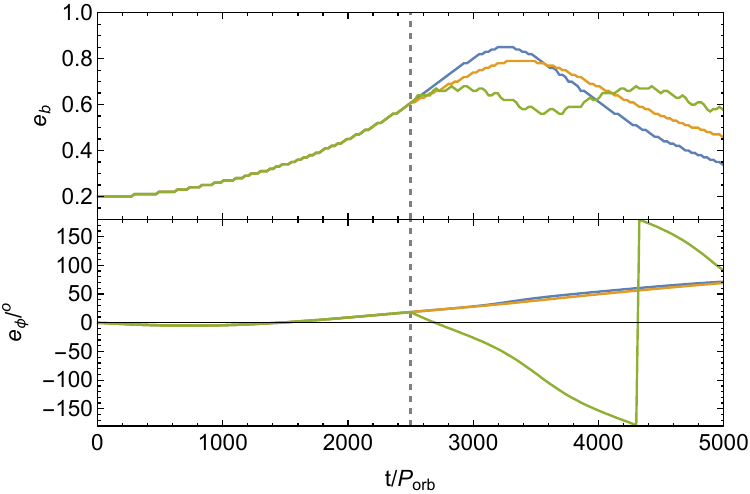}
	\end{center}
    \caption{Binary eccentricity and $e_\phi$ for the $n$-body simulations.  The blue lines here show the three-body simulation that is the same as the blue lines of Fig.~\ref{ecc}. This simulation has $M_1=M_2=30\,\rm M_\odot$, $M_3=0.033\,M_{\rm b}$, $a_3=7.8\,a_{\rm b}$, $i=130^\circ$ and $\Omega=90^\circ$ initially. The yellow lines show a simulation in which the mass of the third body decreases at time $t=2500\,P_{\rm orb}$ to $0.68\,M_3$  and its semi-major axis moves to $7.7\,a_{\rm b}$. The green lines show the four-body simulation in which the outer particle is replaced by two particles to represent the two parts of a broken disc, the inner polar disc and the outer misaligned disc. The particles have masses of $0.22\,M_3$ and $0.68\,M_3$  and semi-major axes of $a_3=3.6\,a_{\rm b}$ and $a_4=7.7\,a_{\rm b}$, respectively.
     }
    \label{ecc2}
\end{figure}

In order to understand the mechanism more clearly, we consider now  some $n$-body simulations in Fig.~\ref{ecc2}. The blue lines show the same three-body simulation as described in Section~\ref{three} and shown in Fig.~\ref{ecc}. The green lines show a simulation in which we replace the outer particle with two particles to represent the inner and outer disc close to  the time when the disc breaks, $t=2500\,P_{\rm orb}$.  In the disc simulation, at $t=2700\,P_{\rm orb}$, the mass of the inner disc is $0.22\,M_{\rm d0}$ and the mass of the outer disc is $0.68\,M_{\rm d0}$, where $M_{\rm d0}$ is the initial disc mass. The angular momentum ratio of the inner disc to the outer disc is 0.22. In the four-body simulation,  the two particles have the same masses as the two parts of the disc and the same angular momentum ratio. The inner particle is at an orbital radius of $3.6\,a_{\rm b}$ and the outer is at $7.7\,a_{\rm b}$.  The inner particle begins at $i=\Omega=90^\circ$ and the outer particle has the same properties as the three-body particle at time $t=2500\,P_{\rm orb}$ except the mass and semi-major axis are slightly smaller.
Fig.~\ref{ecc2} shows the binary evolution. The evolution of the binary eccentricity and $e_\phi$ are remarkably similar to the disc simulation. The eccentricity oscillations become much smaller in amplitude with the polar inner particle. The apsidal precession also becomes much faster and in a retrograde direction. Finally, the yellow lines show a three-body simulation but with the outer particle mass set to $0.68\,M_{\rm d0}$ and orbital radius of $7.7\,a_{\rm b}$. This shows that the effect of the decreasing disc mass is not so significant as the formation of the inner polar ring. 

\section{Extension of the parameter space}
\label{massive}

We now consider the applicability of the mechanism to varying disc mass, black hole masses and disc parameters. In order to explore a wide range of parameter space that is not feasible with circumbinary disc simulations, we use just the three-body simulations to explore the effect of changing the mass of the third body and  varying black hole masses. 

\subsection{Mass of the companion object}

\begin{figure}
    \centering
    \vspace{-0.5cm}
    \includegraphics[width=\columnwidth]{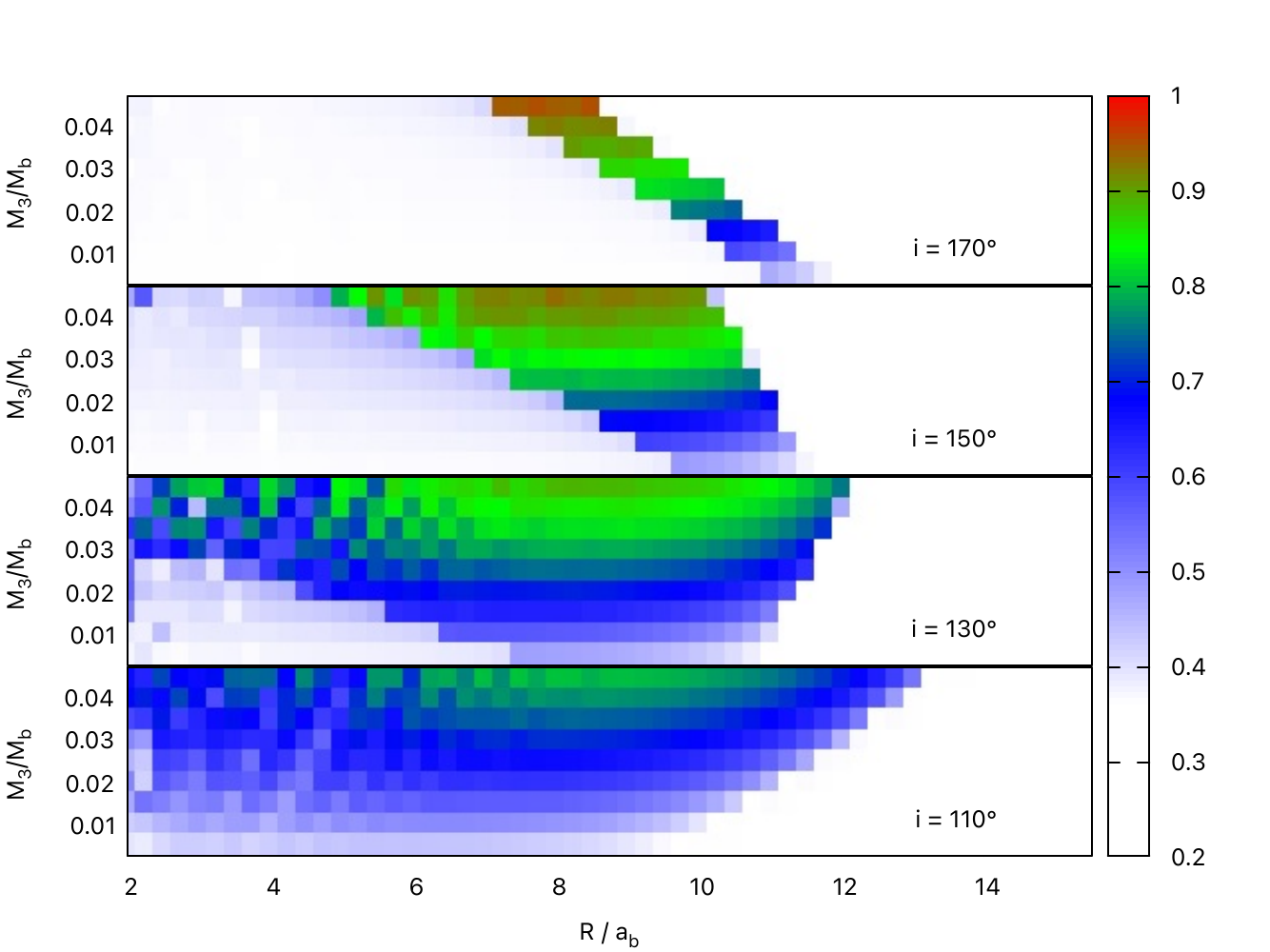}
    \caption{ The colours show the maximum eccentricity of the inner binary for a three-body system for varying companion mass, $M_3$, and orbital radius, $R$, for four different initial inclinations. The inner binary is equal mass with $M_{\rm b}=60\,\rm M_\odot$, $a_{\rm b}=10\,\rm R_\odot$ and $e_{\rm b}=0.2$ initially.  }
    \label{m3}
\end{figure}

 Fig.~\ref{m3} shows the maximum eccentricity of the binary in the three-body simulations in which we vary the mass, orbital radius and inclination  of the third body. The largest eccentricities are reached for a third body that is inclined by $170^\circ$ (top panel), however, the radial range for the orbit is narrow. A disc is unable to drive such high eccentricity growth beginning at this inclination because it radially spreads on a shorter timescale than the ZKL oscillation can be driven. 
For lower initial disc inclination, the radial range of orbits that can drive ZKL of the binary increases. Therefore ZKL driving from a disc is easier at $i=130^\circ$. This is also the most effective inclination since for lower inclinations the maximum eccentricity decreases.

\subsection{Mass of the black hole binary}

Fig.~\ref{m3} shows that there is maximum radius for which the particle can drive binary eccentricity growth. Material inside of this radius  can drive the growth. If there is any material outside of this radius, it is precessing too slowly to keep up with the binary and is found in circulating orbits \citep{Childs2023b}. We can estimate analytically the radius at which the material needs to be accreted for this mechanism to work. \cite{Lepp2023b} found the critical radius at which the stationary inclination  reaches  $180^\circ$ to be
\begin{equation}
     \frac{r_{\rm c1}}{a_{\rm b}} = \left[\frac{ \left(
     \frac{3\Omega_{\rm b}}{4}\right) 
    \left(\frac{M_1M_2}{M_{\rm b}^2}\right) (1+4e_{\rm b}^2)}{-\dot\omega_{\rm GR}-\dot\varpi_{\rm b, \rm quad}}\right]^{2/7},
    \label{rc2}
\end{equation}
where the angular frequency of the binary is $\Omega_{\rm b}=\sqrt{G M_{\rm b}/a_{\rm b}^3}$. This is the radius outside of which there are no librating orbits.
The apsidal precession of the binary driven by GR is
\begin{equation}
    \dot \omega_{\rm GR}=\frac{3(GM_{\rm b})^{3/2}}{a_{\rm b}^{5/2}c^2(1-e_{\rm b}^2)}
\end{equation}  
and the precession rate for the longitude of the periastron of the binary caused by the small mass companion is given by 
\begin{equation}
    \dot \varpi_{\rm b,\rm quad}  = \frac{3}{4}\Omega_{\rm b}\frac{M_3}{M_{\rm b}}\left(\frac{a_{\rm b}}{R}\right)^3 F_2.
\end{equation}
A more general expression for $F_2$ is given in \cite{Lepp2023b} but at an inclination of $180^\circ$ it  simplifies to 
\begin{equation}
F_2=
(1-e_{\rm b}^2)^{1/2} 
\end{equation}
\citep[][]{Zanardi2018,Zhang2019,Lepp2023b}.  For the initial parameters of the disc simulation, we have $r_{\rm c1}/a_{\rm b}=10.3$. However, during the simulation the binary eccentricity evolves, as does the surface density profile of the disc.  

We now examine how the critical radius scales with mass and length.  For a fixed binary eccentricity, we have $\dot \omega_{\rm GR} \propto \sqrt{M_{\rm b}^3/a_{\rm b}^5}$. We take $M_3\propto M_{\rm b}$, $R \propto a_{\rm b}$ and find  $\dot \varpi_{\rm b,\rm quad} \propto \sqrt{M_{\rm b}/a_{\rm b}^3}$. The numerator of the fraction in equation~(\ref{rc2}) also scales with $\sqrt{M_{\rm b}/a_{\rm b}^3}$.
Since the event horizon of  a black hole scales linearly with mass it is reasonable to consider the case where both mass and length share the same overall scaling, so $M_{\rm b}\propto a_{\rm b}$. In this case all of the terms scale as $1/M_{\rm b}$ and $r_{\rm c}/a_{\rm b}$ is scale free.  It should be noted, the same is true for the other critical radii in \cite{Lepp2023b}. So long as the masses and lengths are scaled by the same amount, the critical radii relative to $a_{\rm b}$ remains unchanged. This was also seen in \cite{Childs2023b} for the test particle case in which the apsidal precession is driven only by GR.

Thus the results described in this work can be scaled to different mass black hole binaries by scaling the semi-major axis by the same factor.  For example, if we considered a binary of two $3\times 10^6\,\rm M_\odot$ black holes,  this is $10^5$ times the mass that  we have considered. If the binary separation is increased by the same factor of $10^5$, so that $a_{\rm b}=10^6\,\rm R_\odot=0.02\,\rm parsec$, then the critical radius relative to the binary separation would be the same.   Although our analytical scaling only works at $90^\circ$ and $180^\circ$ we have run numerical tests that suggest that this scaling works for this system.

\subsection{Disc lifetime}

The considerations in the previous two subsections have not taken into account the lifetime of the disc. For this mechanism to operate requires the disc lifetime to be much longer than the ZKL oscillation timescale. The lifetime of the disc is approximately  the viscous timescale that is given by
\begin{equation}
    \frac{t_\nu}{P_{\rm orb}} =\frac{1}{2 \pi \alpha (H/R)^2}\left(\frac{R}{a_{\rm b}}\right)^{3/2}
\end{equation}
\citep[e.g.][]{Pringle1981}.  Note that $t_{\nu}/P_{\rm orb}$ is independent of the binary mass. An approximate ZKL timescale can be found with 
\begin{equation}
    \frac{t_{\rm ZKL}}{P_{\rm orb}}\approx \frac{M_{\rm b}}{M_{\rm d}}\left(\frac{R}{a_{\rm b}}\right)^3
\end{equation}
for a circular orbit outer body \citep{Ford2000,Kiseleva1998}, where $M_{\rm d}$ is the disc mass. This suggests that for a disc mass that is a fixed fraction of the binary mass ($M_{\rm d}\propto M_{\rm b}$),  the ZKL timescale is independent of the binary mass. Note that we require $t_\nu \gg t_{\rm ZKL}$ since as the disc mass decreases, the ZKL effect weakens.

For the parameters of our simulation, with $R=7.8\,a_{\rm b}$, we have $t_\nu /P_{\rm orb}\approx 3.5\times 10^5$, much longer than the ZKL oscillation timescale that is a few thousand binary orbits (see Fig.~\ref{ecc}). 
For a disc with a larger disc aspect ratio, $H/R=0.05$, the viscous timescale can become comparable to the ZKL oscillation timescale within a factor of a few. In this case, the disc is depleted before the binary eccentricity grows, as discussed in Section~\ref{sec:cbd}. In general, we expect the mechanism to work for a range of black hole masses provided that the disc mass is scaled accordingly and that the disc aspect ratio is sufficiently small.

\section{Conclusions}
\label{concs}

We have found that a highly misaligned disc around a binary black hole can significantly increase the eccentricity of a black hole binary through the ZKL mechanism. Disc breaking that occurs during the high eccentricity phase can lead to an inner polar ring. The polar ring drives fast retrograde apsidal precession of the binary orbit which suppresses the ZKL effect and the binary eccentricity can remain high. Therefore, the capture of retrograde circumbinary material can speed up the merger more than the capture of a third body. This has significant implications for the merger timescale for black hole binaries that accrete material through chaotic accretion processes.

\section*{Acknowledgements}
We thank the referee, Hossam Aly, for providing useful comments. Computer support was provided by UNLV’s National Supercomputing Center. We acknowledges support from NASA through grants 80NSSC21K0395 and 80NSSC23M0104.  ACC acknowledges support from the NSF through grant NSF AST-2107738. CJN acknowledges support from the Science and Technology Facilities Council (grant number ST/Y000544/1), and the Leverhulme Trust (grant number RPG-2021-380). We acknowledge the use of SPLASH \citep{Price2007} for the rendering of Fig.~\ref{splash}.

\section*{Data Availability}

 The data underlying this article will be shared on reasonable request to the corresponding author.
 



\bibliographystyle{mnras}
\bibliography{mainmnras} 








\bsp	
\label{lastpage}
\end{document}